\documentstyle[psfig,prd,aps]{revtex}

\begin{document}

\def\kk{{\bf k_1}}
\def\kg{{\bf k_2}}
\def\qq{{\bf q}}
\def\pom{{I\!\!P}}

\draft
\title{Unitarity Corrections to the Structure Functions through the Dipole
Picture}
\author{
 M. B. Gay Ducati
$^{\star}$\footnotetext{$^{\star}$E-mail:gay@if.ufrgs.br} and  M. V. T.
Machado $^{\star \star}$\footnotetext{$^{\star
\star}$E-mail:magnus@if.ufrgs.br}  }  
\address{ Instituto de F\'{\i}sica, Universidade
Federal do Rio Grande do Sul\\ Caixa Postal 15051, CEP 91501-970, Porto
Alegre, RS, BRAZIL}

\maketitle
\begin{abstract}
We study the dipole picture for the description of deep inelastic
scattering, focusing on the structure functions which are driven directly by
the gluon distribution. 
 One performs estimates using the effective dipole cross
section given by the Glauber-Mueller approach in QCD, which encodes the
corrections due to the unitarity effects associated with the saturation
phenomenon. We also address issues about frame invariance of the calculations
when analysing the observables.
 \end{abstract}


\bigskip


\section{Introduction}

The scattering experiments on deep inelastic electron-proton (DIS)  at HERA
have provided measurements of the inclusive structure function $F_2(x,Q^2)$ as
well as of the $F_L$ and $F_2^{c\bar{c}}$ in very small Bjorken variable $x$
values ($x \ll 10^{-2}$). In these collisions the proton target is analysed
through a hard probe with virtuality $Q^2=-q^2$. The momentum fraction is   $x
\sim Q^2/2p.q$, where  $p$ and $q$ are the four-momenta of the incoming proton
and of the virtual photon probe. In the  kinematical region of small $x$, the
gluon is the leading parton driving the  behavior of the deep inelastic
observables. The standard QCD evolution \cite{DGLAP} gives a powerlike growth
for the gluon distribution and related quantities, and this feature leads, in
principle, to the unitarity violation at asymptotic energies, requiring  a
control of the gluon distribution in high energies. In the partonic language,
in the infinite momentum frame,  the small momentum fraction region
corresponds to the high parton density domain, which  is connected
with the black disk limit of the proton target and with the parton
recombination phenomenon. These issues can be addressed through a non-linear
dynamics beyond the usual DGLAP formalism (for a review, see
\cite{GayDucati}). The complete knowledge about the non-linear dynamical
regime plays an important role in the theoretical description of the reactions
in the forthcoming experiments RHIC and  LHC.

In the Breit frame, the QCD factorization theorem allows to calculate
the scattering processes through the convolution of the partonic subprocess
with the parton distribution functions (pdf's). There, the degrees of freedom
of the theory are the quasi-free partons (quarks and gluons). On the other
hand, more recently the theoretical description of the small $x$ physics has
been widely  analyzed in the target rest frame, which is a powerful tool
concerning an unified picture for  both the inclusive and the diffractive
scatterings, including vector meson production \cite{Amirim}. Now, the degrees
of freedom are the color dipoles, which are the most simple configurations
considering  the virtual photon Fock states expansion.  Its main appeal is a
quite simplified picture for the different mentioned processes,
based on  general properties of quantum mechanics. 

The description of DIS in the color dipole picture is quite intuitive,
allowing a simple representation instead of the  involved one from the Breit
(infinite momentum) frame, and such  framework was proposed by Gribov many
years ago \cite{Gribov}. Considering  small values of the Bjorken variable
$x$, the virtual photon fluctuates into a $q\bar{q}$ pair (dipole) with fixed
transverse separation $r$ at large distances upstream of the target and
interacts in a short time with the proton. More complicated configurations
should be considered for larger transverse size systems, for instance the
photon Fock state $q\bar{q} \,+\,\rm{gluon}$. An immediate consequence of
the lifetime of the pair ($l_c=1/2 m_p x$) to be bigger than the interaction's
one is the factorization between the photon wavefunction and the cross section
dipole-proton in the $\gamma^*\,p$ total cross section. The wavefunctions are
perturbatively calculable, namely  through QED for the $q\bar{q}$
configuration \cite{Nikolaev} and  from  QCD for  $q\bar{q}G$ 
\cite{Wusthoff}. The effective dipole cross section should be
modeled and encloses both perturbative and non-perturbative content.
However, since the interaction strength relies only on the configuration of the
interacting system the dipole cross section turns out to be universal and may
be employed in a wide variety of small $x$ processes \cite{Amirim}.

Currently, there are several models for the dipole cross section based either
in pure phenomenological parametrizations or in a more theoretical ground
(for a  review, see Ref. \cite{Amirim}). The main feature in those
models is the description of the energy dependence of the  interaction,
namely taking into account the interplay between hard and soft domains. The
dipole cross section should be consistent with the sharp growth on energy at
small transverse separation $r$ (large gluon density) and a softer  behavior
for larger $r$ (Regge-like phenomenology). An additional ingredient is the
expectation for saturation effects in high energies as a consequence of  the
unitarity requirements. Indeed, the growth of the gluon distribution should be
tamed at very small $x$ and it has been found that the corrections are
important already in the present HERA kinematics \cite{Levin}. Moreover, such
effects are associated with higher twist contributions concerning the standard
linear evolution equation, i.e. the  DGLAP formalism \cite{BGP}. 

Here, we take into account a formalism  providing 
the unitarity corrections to the deep inelastic scattering at small $x$,
namely the Glauber-Mueller approach in QCD. It was introduced by A.
Mueller \cite{Mueller90}, who developed the Glauber formalism to study
saturation effects in the quark and gluon distributions in the nucleus
considering  the  heavy onium scattering.
Afterwards, the authors of Ref. \cite{Ayala} extended that approach that has
as a limit the  GLR results \cite{GLR}. It is obtained  an evolution
equation taking into account the unitarity corrections (perturbative
shadowing), generating  a non-linear dynamics which is related with higher
twist contributions. Its main characteristic is to provide a theoretical
framework for the saturation  effects, relying on the  multiscattering of the
pQCD Pomeron. The latter is represented through the usual gluonic ladder in the
double logarithmic approximation. 

Summarizing the Glauber-Mueller approach, the gluonic content of the nucleus
or nucleons is obtained in the following way: in the rest frame, a virtual
probe (gluon) decays into a gluon pair  interacting with
the nucleon inside the nucleus. The multiple scatterings of the pair
give rise to the unitarization of the corresponding cross section. The
calculations are performed in the double logarithmic approximation (DLA),
corresponding to the condition $\alpha_s \ll \gamma_G \ll 1$, where $\alpha_s$
and $\gamma_G$ are the QCD coupling constant and the gluon anomalous
dimension, respectively. In this approximation, the transverse separation of
the pair remains fixed allowing an eikonal description for the interaction
gluon pair-nucleon through incoherent  multiscatterings \cite{Ayala}. The
cross section for the interaction can be expressed in terms of the nucleon
gluon distribution $xG(x,\,\tilde{Q}^2)$ and of the transverse separation $r$
of the gluon pair. The procedure for an initial state  quark-antiquark pair is
similar, up to proper color coefficients. Such a formulation has produced
comprehensive phenomenological applications:  the inclusive structure function 
$F_2$ \cite{Ayalaepjc}, the longitudinal $F_L$ and charmed  $F_2^{c\bar{c}}$
ones \cite{PRD59}, the logarithmic slope $\partial F_2/\partial \log Q^2$
\cite{Victorslope} and other related quantities have been calculated in
the Breit system. The respective nuclear case has been investigated in Refs.
\cite{Victornuclear} and the equivalence with other high density QCD
approaches has been reported in Refs. \cite{VicHDQCD}. Moreover, the
asymptotic limit of the inclusive structure function and its logarithmic slope
are estimated in \cite{Unitbound}.  Regarding the rest frame, Glauber-Mueller
has been also used to estimate the saturation effects for  DIS and diffractive
dissociation \cite{Lev1} as well as it is considered as initial condition for
a high energy evolution equation (for a review, see \cite{Levin}).
Therefore, the Glauber-Mueller approach gives a good  framework for the
unitarity effects (saturation) in the nucleon and nuclear sectors,
providing the dynamics of the observables in a quantitative level.

In this work we make use of the parton saturation formalism to study the
description of the observables driven by the gluonic content of the proton in
the color dipole picture. The inclusive structure function $F_2$ is
calculated properly, disregarding  approximations commonly
considered in previous calculations \cite{Ayalaepjc,Victorslope,Unitbound}.
The structure functions $F_L$ and $F_2^{c\bar{c}}$ are presented for the first
time using the Glauber-Mueller approach and the rest frame in comparison with
the experimental data.  The saturation effects are included in the effective
dipole cross section corresponding to the small size dipole contributions. The
large dipole sizes are taken into account through an ans\"{a}tz for the
non-perturbative region. Here, we choose to freeze the gluon distribution at
large distances \cite{Lev1}. The dipole framework provides a clear
identification in the transverse distance $r$ range where the perturbative and
non-perturbative sectors (soft domain) contribute to the calculated 
quantities. Large dipoles correspond to soft domain and the  small ones are
connected with the hard sector. As we will see, the photon wavefunctions play
the role of a weight function selecting small dipole sizes, with a 
non-negligible contribution coming from the large dipole
sizes. However, when considering the production of heavy
quark pairs (charm and bottom), their masses  largely diminish the soft
contribution.  These issues are addressed throughout this paper.

This work is organized as follows. In the next section we shortly review
the deep inelastic scattering in the rest frame, introducing  the main formulae
for the further analysis. In Sec. (3) one addresses the effective
dipole cross section considering saturation effects (unitarity corrections)
encoded in the Glauber-Mueller approach, pointing out its main properties and
discussing the large transverse separation contribution. The Sec. (4) is
devoted to calculate the theoretical estimates for the HERA observables,
focusing on those ones dominated by the gluon content. In the last
section we draw our conclusions and comments.

\section{The Deep Inelastic Scattering in the Proton Rest Frame}
\label{Sec2}

In the reference frame where the target (proton) has infinite momentum the
usual description of the dynamics evolution is the following: the emitted
partons from the target remain quasi-free for enough time  in such a way that
the  virtual probe (photon) detects them as real particles (asymptotic
freedom). On the other hand, in the small $x$ region it is suitable to
describe the evolution in a system where the target is at rest: in this
situation the evolution is related to the partonic fluctuations of the probe
and their interactions with the nucleon. The rest frame physical
picture  is advantageous since the lifetimes of the photon fluctuation and
of the interaction process are well defined \cite{BrodDelduca}. The simplest
case  is the quark-antiquark state (color dipole), which is the
leading configuration for  small transverse size systems. Its life time
can be estimated by the uncertainty principle through the energy fluctuation
associated with the emerging pair. The well known coherence length is
expressed as $l_c=1/(2xm_p)$, where $x$ is the Bjorken variable and $m_p$ the
proton mass. For instance, in deep inelastic at HERA kinematics reaching at $
x\sim 10^{-5}$, the coherence length is about $10^4$ fm, which  is a
distance larger than the radius of any atomic nuclei. An immediate consequence
of such a  picture is the factorization between virtual photon wavefunctions
and the  interaction cross section on the corresponding amplitude in the impact
parameter space representation.

\begin{figure}[t]
\centerline{\psfig{figure=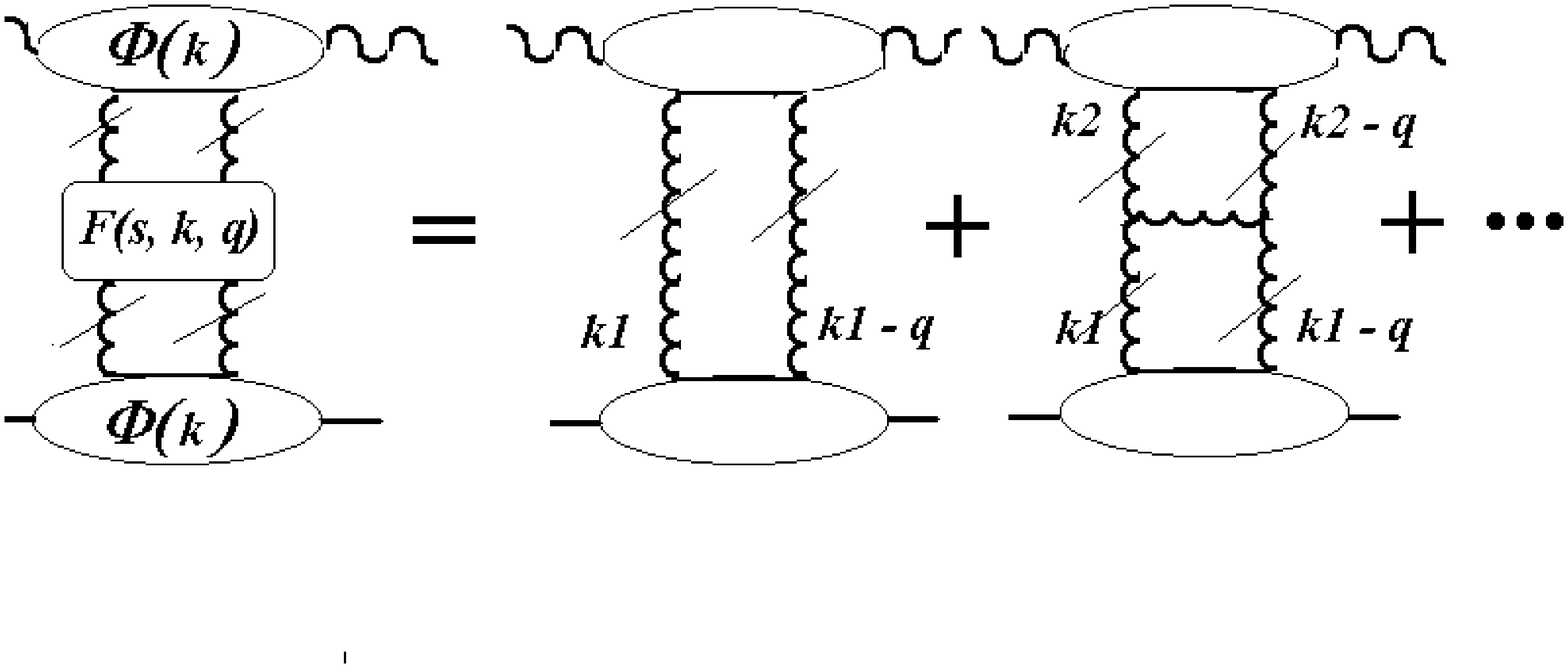,width=9cm}}       
\caption{Representation of the deep inelastic process, where the upper blob
corresponds to the photon impact factor and the bottom one represents the
proton impact factor. The corresponding two first orders in perturbative
expansion (BFKL-like) are depicted.}  
\end{figure}

A striking consequence of the formulation above is that the photoabsortion
cross section can be derived from the expectation value of the interaction
cross section  for the multiparticle Fock states of the virtual
photon weighted by the light-cone wave functions of these states
\cite{Nikolaev}. The scattering matrix has an exact diagonalization in the
$\gamma^*$ Fock states representation, where the partial-wave amplitudes
subjected to the $s$-channel unitarization are identified properly. Then, the
photoabsortion cross section can be cast into the quantum mechanical
factorized  form, \begin{eqnarray}
\sigma^{\gamma^*p}_{T,L}(x,\,Q^2)=\int \,d^2{\bf r}\,\int_0^1\,dz \,\,|\Psi
_{T,L}(z ,\,{\bf r})|^2\,\,\sigma^{\rm{dipole}}(x,z,{\bf r})\,\,, \label{eq8}
\end{eqnarray}

The formulation above is valid even beyond perturbation theory, since it
is determined from the space-time structure of the process. The $\Psi_{T,L}(z
,\,{\bf r})$ are the photon wavefunctions (for transverse, $T$, and
longitudinal, $L$, polarizations)  describing the pair configuration, where $z$
and $1-z$ are the fraction of the photon's light-cone momentum  carried by the
quark and antiquark of the pair, respectively. The transverse separation of
the pair is $\bf{r}$. The precise normalizations of the wavefunctions can be
determined through the Fock expansion
$|\gamma_{\rm{phys}}\!>=\,\sqrt{Z_3}\,|\gamma_{\rm{bare}}\!> + \,{\cal
N}_{q\bar{q}} \,|q\bar{q}\!>$, where $Z_3$ is the $\gamma$-wavefunction
renormalization constant, $|\gamma_{\rm{bare}}\!> $ denotes the bare state and
${\cal N}_{q\bar{q}}$ is the coefficient determining the probability of the
$q\bar{q}$ pair fluctuation  in the photon \cite{Forshaw}. Considering
completely normalized states, then ${\cal N}^2_{q\bar{q}}=1-Z_3$ and the
normalization for transverse photons is obtained from $\int \,dz\,d^2{\bf r}
\,\,|\Psi _{T,L}(z ,\,{\bf r})|^2= {\cal N}^2_{q\bar{q}}$. The remaining
normalizations, i.e., longitudinal component and cross section, are
consequently fixed \cite{Forshaw}. The explicit expressions are well known,
\begin{eqnarray} \vert \Psi_{T}\,(z,{\bf   r}) \vert^2  \:=\:
\frac{6\, \alpha_{\rm{em}}}{4\,\pi^2 } \: \sum_{i}^{n_f}\,e_i^2 \:
\Big\{\,[\,z^2+(1-z)^2\,]\:\varepsilon^2\,K_1^2\,(\varepsilon\, r) 
\,+\, m_q^2\: K_0^2\,(\varepsilon\, r)\,\Big\} \label{eq9}
\end{eqnarray}
and
\begin{eqnarray}
\vert \Psi_{L}\,(z,{\bf   r}) \vert^2  \:=\:
\frac{6\, \alpha_{\rm{em}}}{4\,\pi^2 } \:\sum_{i}^{n_f}\,e_i^2 \:
\Big\{\,4\,Q^2\,z^2\,(1-z)^2\,K_0^2\,(\epsilon\, r)\,\Big\}\,. \label{eq10}
\end{eqnarray}

Clarifying the notation, the auxiliary variable
$\varepsilon^2=z(1-z)Q^2 + m_q^2$, with $m_q$
the light quark mass, and   $K_0$ and $K_1$ are the  Mc Donald
functions of rank zero and one, respectively. 

The quantity $\sigma^{\rm{dipole}}(x,z,{\bf r})$ is interpreted as the cross
section of the scattering of the effective dipole with fixed transverse
separation ${\bf r}$ \cite{Nikolaev}. It is directly dependent on the
unintegrated gluon distribution,
\begin{eqnarray}
\sigma^{\rm{dipole}}(x,z,{\bf r})= \frac{4\pi
\alpha_s}{3}\,\int \frac{d^2 \kk}{\kk^4}\, {\cal F}(\,\frac{x}{z}, \kk)\, 
 \left( 1- \rm{e}^{i\,\kk .{\bf r}} \right) \label{eq11}
\end{eqnarray}

The unintegrated gluon distribution, ${\cal F}(\,\frac{x}{z}, \kk)$, vanishes
at the gluon transverse momentum $|{\bf k}_1| \rightarrow 0$ (and similarly for
the factor $( 1- \rm{e}^{i\,\kk .{\bf r}})$) due to gauge invariance, and
therefore the dipole cross section has to be   infrared finite. The
most important feature of the dipole cross section is its universal character,
namely it depends only on the transverse separation ${\bf r}$ of the color
dipole. The dependence on the external probe,  i.e. the photon
virtuality, is included in the wavefunctions. 

The main technical difficulty in Eq. (\ref{eq11}) is to model the unintegrated
gluon distribution function in a suitable way, mainly in the small transverse
momentum ($k_T$) region (infrared sector). In particular, to obtain these
distributions one should solve numerically an evolution equation,
which makes the procedure cumbersome for practical use (instead, for a prompt
parametrization see \cite{Ivanov}).  In general, this  is avoided
introducing  an ans\"{a}tz for the effective dipole cross section and
analyzing the process in the impact parameter space.  The
main feature of the current models in the literature is to interpolate the
physical regions of small transverse separations (QCD-parton improved
model picture) and the large ones (Regge-soft picture). Below, we quote two of
them, which have connections with the work performed here.

The  phenomenological saturation model of
Golec-Biernat and W\"{u}sthoff (hereafter GBW) gives a good description of DIS
data \cite{Golec}. The corresponding cross section interpolates between color
transparency, i.e. $ \sigma^{\rm{dipole}}\sim r^2$ at small $r$,  and constant
cross section $\sigma_0$ at large $r$ (confinement). Such a procedure ensures
$Q^2$-saturation, while parton  saturation  at low $x$ is obtained with an
eikonal-inspired shape for the dipole cross section, \begin{eqnarray}
\sigma^{\rm{GW}}(x,r^2)& = & \sigma_0 \left(1-e^{\,-r^2/R^2_{\rm{sat}}(x)}
\right) \,,\label{eq13} \\ R^2_{\rm{sat}}(x) & = &
4\,R_0^2(x)=\frac{4}{Q^2_0}\left( \frac{x}{x_0} \right)^{\lambda}\,, \nonumber 
\end{eqnarray}
where the parameters $\sigma_0=23.03$ mb, $x_0=3.04\,10^{-4}$ and
$\lambda=0.288$ are fitted to the HERA DIS inclusive data with~ $x<10^{-2}$,
whereas  $Q^2_0=1$ GeV$^2$ sets the dimension. The $x$-dependent saturation
radius $R_0^2(x)$ scales the pair separation $r$ in the cross section and is
associated with the mean separation between partons in the nucleon. This
approach was used to describe diffractive dissociation in a parameter-free
way, considering also the required $q\bar{q}G$ configuration. Some criticism
to the GBW model, mainly concerning a better knowledge of the parton (gluon)
distribution at low $r$ and its saturation at  small $x$ are postponed to the
next section. The most of them  can also to be found in Refs.
\cite{Amirim,Lev1}.

Our work in this paper is closer to the McDermott et al. one \cite{McDermott}
(hereafter McDFGS), where perturbative QCD relates the dipole cross section to
the leading logarithmic gluon distribution function in the proton, at
LLA($Q^2$) accuracy, 
\begin{eqnarray}
\sigma^{\rm{McDFGS}}= \frac{\pi^2
\alpha_s(\tilde{Q}^2)}{3}\,r^2\, x\,G(x,\tilde{Q}^2) \,,
\label{eq14}
\end{eqnarray} 
where the identification $\tilde{Q}^2 \approx Q^2$ is allowed at leading-log
level. A study of the dipole  cross section is carried for all transverse
separations $r$ in \cite{McDermott}. The small $r$ region is described by the
usual gluon pdfs, evoluting with the scale $\tilde{Q}^2=10/r^2$, while for
large separations ($r_{\rm{pion}} \geq 0.65$ fm) the dipole cross section is
driven by the  pion-proton contribution with the typical soft energy behavior
from the hadronic sector. Moreover, the taming of the parton density is
implemented by hand starting at the named critical transverse separation
$r_{\rm{crit}}$, stated when the dipole cross section reaches one half of its
maximum value labeled by the pion-proton cross section. In connection with the
present work, the Glauber-Mueller approach gives  the Eq.
(\ref{eq14}) at the  Born level. Instead of  using an ad hoc control of the
gluon distribution, the Glauber-Mueller provides corrections required by
unitarity in an eikonal expansion. For the large $r$ region, we choose to
follow a similar procedure from the  GBW model, namely saturating
 the dipole cross section ($r$-independent
constant value).

Having defined the notation and reviewed the main properties settled by the
rest frame representation of deep inelastic process, in the next section we
address the unitarity corrections formalism contained in the
Glauber-Mueller approach, that we will consider in the calculation of $F_2$,
$F_L$ and $F_2^{c\bar{c}}$.

\section{The Glauber-Mueller Approach}

The Glauber formalism concerns mainly  interactions with a target nucleus, allowing to  calculate the amount of unitarity corrections to the
nuclear cross section. However this approach can be extended to take into
account the evolution of the partonic densities (saturation) through the
multiple scatterings. Below, we review shortly the main properties of the
Glauber formalism in QCD, either in the nuclear case as its application to the
nucleon case. We indicate the original papers \cite{Ayala} for a complete
presentation. It should be stressed that  saturation
effects mean unitarity corrections to the observables. Indeed,
the asymptotic calculations have produced an unified $\ln(1/x)$ pattern for the
cross section and gluon function instead of a effectively saturated one
\cite{Unitbound}.

Since the small $x$ limit is driven by the gluonic interactions, we consider a
virtual probe $G^*$ with invariant mass $Q^2$ which decays into a gluon pair
$GG$ having a transverse separation ${\bf r}$ and transverse momentum ${\bf
k}$. Then, the pair interacts with the target through a gluonic ladder at
fixed transverse separation. Following the discussion from the previous
section, the photoabsortion cross section for the probe particle in terms of
$x$ and virtuality $Q^2$ in the nuclear case is
\begin{eqnarray}
\sigma^{G^*\,A}_{tot}(x,\,Q^2)=\int \,d^2{\bf r}\,\int_0^1\,dz \,\,|\Psi
_{GG}(z ,\,{\bf r}, Q^2)|^2\,\,\sigma^{GG-\rm{nucleus}}(x,z,{\bf r})\,,
\label{glauber}
\end{eqnarray}
where the variables have the same identification as in the $q\bar{q}$ pair
discussion. The quantity $\Psi _{GG}$ is the light-cone wavefunction for the
gluon pair. The Glauber's multiscattering theory employs the phase shift
method to describe processes at high energy for an incident particle
undergoing into  successive scatterings. When $x$ is small, the coherence
length $l_c$ is bigger than the mean radius $R_A$ for all nuclei and the
interaction of the initial parton stands through the entire nuclear path, 
providing coherent interactions with all target partons along the distance
$l_c$. These scatterings are coherent in an interaction of a hadronic
fluctuaction, i.e. quark or gluon pairs, with the nucleus. There are
interference effects among them,  generating a reduction in the nuclear cross
section, $\sigma^{\rm{nucleus}}< A\,\sigma^{\rm{nucleon}}$. Otherwise, when
$l_c\leq R_A$,  full incoherent scatterings occur leading to the  expectation
that the  nuclear cross section equals to $A\,\sigma^{\rm{nucleon}}$. The well
known Glauber formula for the total cross section of a hadronic state with the
nucleus is  \begin{eqnarray} \sigma_{tot}^{\rm{nucleus}}=2\,\int d^2{\bf b}\,
\left(1-\rm{e}^{-\frac{1}{2}\,\sigma_{\rm{nucleon}}\,S_A({\bf b})} \right)\,,
\label{glauber2} \end{eqnarray} where $S_A({\bf b})$ is a profile function
containing the dependence on the impact parameter ${\bf b}$, which is the
conjugate variable to the momentum transfer $t$. It is related to the nucleon
distribution inside the nucleus and  encodes the information about the angular
distribution of the scattering. We discuss its particular  shape later on. The
Eq. (\ref{glauber}) is quite general and allows to describe the hadronic
fluctuations of the photon or any virtual probe, as saw in the last section.

Considering the scattering
amplitude dependent on the usual Mandelstan variables $s$ and $t$, now written
in the impact parameter representation ${\bf b}$,   \begin{eqnarray}
a(s,{\bf b})\equiv \frac{1}{2\pi}\, \int\,d^2{\bf q}\, \rm{e}^{-i\,{\bf
q}.{\bf b}}\, {\cal A}\,(s,t=-q^2)\,.
 \end{eqnarray}
the corresponding total and elastic cross sections (from
Optical theorem) are rewritten in the impact parameter representation (${\bf
b}$) as 
\begin{eqnarray}
\sigma_{tot} & = & 4\,\pi\,Im {\cal A}\,(s,0)=2\,\int \,d^2{\bf b}\,Im \,\,
a(s,{\bf b})\,,\\ \sigma_{el} & = & \int \,d^2{\bf b}\, |a(s,{\bf b})|^2\,,
\end{eqnarray}
A very important property when treating the scattering in the impact
parameter space is  the simple definition for the unitarity constraint
\cite{Ayala}. At fixed ${\bf b}$, the constraint can be expressed in the
following way, \begin{eqnarray}
\sigma_{tot} & = &  \sigma_{el}+ \sigma_{inel}\,,\\
2\,Im\,a(s, {\bf b}) & = & |a(s,{\bf b})\,|^2 + C_{in}(s,{\bf b})\,,
\end{eqnarray}
with $C_{in}(s,{\bf b})$ denoting the sum of contributions  from  all
the inelastic channels. The constraint above has a simple solution. If the real
part of the scattering amplitude vanishes at the high energy limit,
corresponding to small $x$ values, the solution is
\begin{eqnarray}
a(s,{\bf b}) & = & i\,\left[ \,1-\rm{e}^{-\frac{1}{2}\,\Omega\,(s,\,{\bf b})}
\, \right]\,,\\ \sigma_{tot} & = & 2\,\int d^2{\bf b}\,
\left[\, 1-\rm{e}^{-\frac{1}{2}\,\Omega\,(s,\,{\bf b})}\,
\, \right]\,, \label{sigeik} \end{eqnarray}
where the opacity $\Omega$ is an arbitrary real function and it should
be determined by a detailed model for the interaction. The opacity function 
has a simple physical interpretation, namely $\rm{e}^{-\Omega}$ corresponds to
the probability that no inelastic scatterings with the target occur. To
provide the connection with  the Glauber formalism, the opacity function
can be written in  the factorized form $\Omega (s,{\bf b})={\Omega}(s)\,S({\bf
b})$, considering  $S({\bf b})$ normalized as $\int d^2{\bf b}\, S({\bf b})=1$ 
(for a detailed discussion, see i.e. \cite{LevNPB}).

From Eqs. (\ref{sigeik}) and (\ref{glauber2}), we
identify the opacity ${\Omega}(s \approx Q^2/x; {\bf
r})=\sigma^{\rm{nucleon}}(x,{\bf r})$. Moreover, it has been found that the
same formalism for multiple scatterings can be applied to the nucleon case.
To proceed, we should determine the $GG$ cross section. The gluon
pair cross section is equivalent to the quark pair one, up to a color factor
($\sigma^{GG}=\frac{9}{4}\sigma^{q\bar{q}}$). The ($q\bar{q}$ pair)
dipole-proton  cross section is  well known \cite{Ayala,LevNPB}, 
calculated starting from Eq. (\ref{eq11}),  and in double logarithmic
approximation (DLA) has the following form  \begin{eqnarray}
\sigma^{q\bar{q}}_{\rm{nucleon}}(x,r)= \frac{\pi^2
\alpha_s(\tilde{Q}^2)}{3}\,r^2\, x\,G(x,\tilde{Q}^2)  \label{dipdla}
\end{eqnarray}
with the $r$-dependent scale $\tilde{Q}^2=r_0^2/r^2$. Considering Eq.
(\ref{dipdla}) one can connect directly the dipole picture with the usual
parton distributions (gluon), since they are solutions of the DGLAP equations.
In our case, we follow the  calculations in Ref. \cite{Ayala,LevNPB} and
consider the effective scale $\tilde{Q}^2=4/r^2$. Such a value differs from
\cite{McDermott}, where it is  $r_0^2=10$, which  is obtained by an averaging
procedure on the transverse size integral of $F_L$. However, in further
studies in vector mesons it was found that $r_0^2$ ranges from 4-15, and $F_2$
and $F_L$ are not sensitive to those  variations. Thus, these values are
consistent in leading logarithmic $Q^2$  approximation.

From the above expression, one obtains a dipole cross section satisfying the
unitarity constraint and a framework to study the unitarity effects
(saturation) in the gluon DGLAP distribution function. Hence, hereafter we use
the Glauber-Mueller dipole cross section given by
\begin{eqnarray}
\sigma_{\rm{dipole}}^{GM}=2\,\int d^2{\bf b}\,
\left(1-\rm{e}^{-\frac{1}{2}\,\sigma^{q\bar{q}}_{\rm{nucleon}}(x,{\bf
r})\,S({\bf b}) }\right)\,. 
\label{dipolgm}
\end{eqnarray}

Here, some comments are in order. The Glauber-Mueller approach is valid in the 
small $x$ region, and the gluon emission is described in the double
logarithmic approximation of  perturbative QCD. The interaction of the quark
pair with the nucleon (proton) occurs through ladder diagrams exchange,
satisfying the DGLAP equations in the  DLA limit. The high energy limit allows
to treat successive scatterings as independent collisions, meaning the
process described by the  eikonal picture of a relativistic particle
crossing the nucleus. Moreover, as a consequence of
no correlation among nucleons inside the nucleus (in the nuclear case),
there are  no correlations among partons from different partonic cascades, 
stressing that only the fastest parton interacts with the target.
The corrections coming from the slowest partons in the cascade (emitted by the
pair) lead to the AGL non-linear evolution equation \cite{Ayala}, and they
have been considered recently to describe diffractive DIS in Ref. \cite{Lev1}.
Regarding criticisms to the Glauber-Mueller approach, we indicate the recent
paper \cite{Kovner} for a complete discussion about the eikonal-like models,
concerning their advantages and limitations as well as pointing out the
improvements to be taken into account to introduce the proper corrections.

Now, we proceed to calculate numerical estimates of the dipole cross section
using the Glauber-Mueller approach through Eqs. (\ref{dipdla}-\ref{dipolgm}).
Then we are calculating saturation effects in the color dipole picture. 
Firstly, we need to discuss the profile function $S( b)$. This function
contains information about the angular  distribution in the scattering, namely
the $t$ dependence (quark pair-ladder and proton-ladder couplings). Both of
them can be approximated by an exponential parametrization, leading to 
a simple gaussian shape in the impact parameter space, $S({b})=\frac{A}{\pi\,R^2_A}e^{\frac{-b^2}{R^2_A}}$, where $A$ is the atomic
number and  $R_A$ is the target radius. We will keep this notation although we
are only concerned with the nucleon case. The $R^2_A$ value should be
determined from data, 
ranging between  $5-10$ GeV$^{-2}$ for the proton case (see discussions in 
\cite{Lev1,Levin}). For nuclear reactions, a more suitable shape for the
profile should be taken into account, since the gaussian approximation is no
longer appropriate to describe the nuclear profile for large $A$. Here, we
have used the value ( $R^2_A=5$ GeV$^{-2}$) obtained from a good description
of both inclusive structure function and its derivative \cite{Victorslope}.
Such a value corresponds to significative unitarity corrections to the standard
DGLAP input even in the current HERA kinematics.

\begin{figure}[t]
\centerline{\psfig{figure=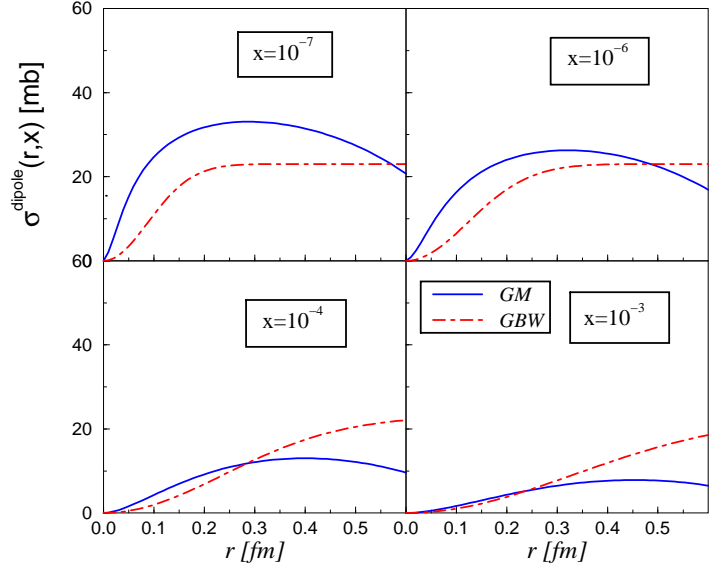,width=9cm}}       
\caption{ The dipole cross section as a function of $r$ at fixed $x$. The
Glauber-Mueller corresponds to the solid line and the GBW model results (dashed
curves) are also included.}
 \label{dipsigmas} 
\end{figure}

Now, we discuss in a detailed way the main characteristics emerging from the
dipole cross section Eq. (\ref{dipolgm}). In order to do this, in Fig.
(\ref{dipsigmas}) one shows the Glauber-Mueller dipole cross section as a
function of dipole transverse size $r$ at fixed momentum
fraction $x$. For  a better illustration on the expected partonic
saturation effects, we run the Bjorken $x$ down to a quite small value
$x=10^{-7}$ (THERA region). Hereafter, we are using the GRV gluon distribution
at leading order \cite{GRV94} in the input Eq. (\ref{dipdla}), whose choice we
justify below. The solid lines correspond to the dipole cross section 
calculation, Eq. (\ref{dipolgm}), whereas the dashed lines are the GBW model
\cite{Golec} presented for comparison.  The general shape in terms of
the dipole size comes from the balancing between the color transparency
$\sigma_{dip} \sim  r^{2}$ behavior and the gluon distribution. 

Here some comments about the large transverse separation are in order:
although perturbative QCD provides  reliable results at small distances
(small dipole sizes), the nonperturbative sector is not still completely
understood. The usual pdf's are evoluted from a perturbative initial scale
$Q_0^2 \approx 1$ GeV$^2$, and one has little information about the behavior
at $Q^2 \leq Q_0^2$. In general one makes use of  Regge phenomenology to
estimate those contributions (see, for instance \cite{McDermott}). Thus,
extrapolating to lower virtuality  regions (large dipole sizes)  one needs a
parametrization regarding the nonperturbative  sector. 

This is the main justification of the use the GRV94
parametrization \cite{GRV94} in our calculations. Bearing in mind that
$Q^2=4/r^2$, its evolution initial scale is $Q_0^2=0.4$ GeV$^2$ allowing 
to scan dipole sizes up to $r_{\rm{cut}}=\frac{2}{Q_0}$ GeV$^{-1}$ (= 0.62 fm).
For the most recent parametrizations, where $Q_0^2 \sim 1$ GeV$^2$
($r_{\rm{cut}} \approx 0.4$ fm) the amount of nonperturbative contribution in
the calculations  should increase. An additional advantage is that GRV94
does not include non-linear effects to the DGLAP evolution  since the
parametrization was obtained from rather large $x$ values. This feature ensures
that the parametrization does not include  unitarity corrections (perturbative
shadowing effects) in the initial scale.

Now, we should introduce an ansatz for the large transverse separation region.
A more phenomenological way is to match the pQCD dipole cross section with the
typical hadronic one $\sigma_{\pi\,N}$ at $r_{\rm{cut}}$ as performed in
\cite{McDermott}. However, due to the significant growth of the pQCD dipole
cross section at high energies, we choose an alternative ansatz: the gluon
distribution is frozen at scale $r_{\rm{cut}}$, namely
$x\,G(x,\,\tilde{Q}^2_{\rm{cut}})$. Then, for the large distance contribution
$r \leq r_{\rm{cut}}$ the gluon distribution reads as 
\begin{eqnarray}
x\,G(x, Q^2 \leq Q_0^2)= \frac{Q^2}{Q^2_0}\,x\,G(x,Q^2=Q_0^2)\,,
\label{ansatz}
\end{eqnarray} 
leading to the correct behavior $x\,G \sim Q^2$ as $Q^2 \rightarrow 0$.
In a more sophisticated case, one can substitute the freezed scale
$\tilde{Q}^2_{\rm{cut}}$ by the saturation scale $Q^2_s(x)$  to take into
account a realistic value of the gluon anomalous dimension in all kinematic
region (see correlated issues in \cite{Lev1}).

Recently, the  phenomenological model of Ref. \cite{Golec} has
produced a good description of HERA data in both inclusive and diffractive
processes. It is constructed interplaying the color transparency behavior
$\sigma_{dip} \sim r^2_{\perp}$ at small dipole sizes and a flat (saturated)
behavior at large dipole sizes $\sigma_{dip} \sim \sigma_0$ (confinement).
The expression has the eikonal-like form,  \begin{eqnarray}
\sigma_{q\bar{q}}(x,r)=\sigma_{0}\left[1-\exp\left(\frac{r^{2}Q_{0}^{2}}
{4(x/x_{0})^\lambda}\right)\right] \,,\end{eqnarray}   
where $Q^2_{0}=1$ GeV$^2$ and
the three fitted parameters are $\sigma_{0}=23.03$ mb, $x_{0}=3.04\,10^{-4}$
and $\lambda=0.288$ and the notation for the saturation radius
$R_0(x)=(x/x_0)^{\lambda/2}$.  The GBW total cross section lies below the
typical hadronic cross section: for instance in the pion-proton case,
convoluting the pion wavefunction squared with the GBW dipole cross section we
would have a constant cross section at high energies $\sigma^{GBW}_{tot} \leq
\sigma^{\pi \,N}_{tot}$. Despite  describing data in good agreement, GBW has
some details that deserve some discussions: the approach does not present a
dynamical hypothesis for the saturation phenomena and does not match DGLAP
evolution. In GBW, saturation is characterized by the $x$-dependent saturation
radius $Q_s^2(x)=1/R^2_0(x)$ instead of the scale coming from Glauber-Mueller,
$\kappa_G(x,Q_s^2)=1$, which can be easily extended for the nuclear case
\cite{Ayala}.

In Fig. (\ref{dipsigmas}) one shows the Glauber-Mueller dipole cross section
and the GBW model \cite{Golec}. We choose to compare them due to the fact that
GBW is actually a particular case of the Glauber-Mueller approach, considering
 an oversimplified profile function. We
should show below that this fact allows  to construct an extended saturation
model with DGLAP evolution \cite{BGK} (BGK model).  In the lower
plots, where $x=10^{-3}$ and $x=10^{-4}$, the GM cross section
underestimates the GBW one. However, as $x$ decreases the gluon distribution
in the proton rises producing a bigger dipole cross section. This
feature is clear in the upper plots, for smaller $x=10^{-6}$ and $x=10^{-7}$,
where GM overestimates GBW by a significant factor mostly at intermediate $r$
values. 

Finally, we show the connection between the Glauber-Mueller approach with the
saturation model with DGLAP evolution \cite{BGK}, observing that it is timely
since the latter is quite efficient in describing data and its qualitative
success corroborates quantitative QCD studies. Considering the particular case
of central collisions, namely scattering at impact parameter $b=0$
($S=A/\pi\,R_A^2$), the Glauber-Mueller approach produces, \begin{eqnarray}
\sigma_{\rm{dipole}}^{GM}(x,{\bf r}, b=0)=2\,\int d^2{\bf b}\,
\left(1-\rm{e}^{-\frac{1}{2}\,\sigma^{q\bar{q}}_{\rm{nucleon}}(x,{\bf
r})\,\frac{1}{\pi\,R^2_A} }\right)\,.  \label{dipolgmb0}
\end{eqnarray}

The integration over $b$ can be promptly carryied out, and introducing the
notation for the proper normalization for the dipole cross section,
$\sigma_0 \equiv 2\,\pi \, \int_{0}^{R^2_A} db^2 = 2\pi\,R_A^2 \,,
\label{normalization}$ the Eq. (\ref{dipolgmb0}) recovers the simple expression for the
saturation model DGLAP evoluted \cite{BGK}, 
\begin{eqnarray}
\sigma_{\rm{dipole}}^{BGK}(x,{\bf r})= \sigma_0\, \left[ \,
1-\exp\left(-\frac{\pi^2 \, {\bf r}^2\,
\alpha_s(\mu^2)\, x G(x,\mu^2)}{3 \,\sigma_0}\right)\right]\,.  \end{eqnarray}

For a phenomenological analysis, the parameter $\sigma_0$ and the
scale $\tilde{Q}^2$ are determined from data in \cite{BGK}. In our case, it
assumes the well defined value $\sigma_0=12.22$ mb, using  $R_A^2=5$
GeV$^{-2}$. For a larger radius, for instance $R^2_A=10$ GeV$^{-2}$, one obtains a value
$\sigma_0=24$ mb,  closer to the GBW one.  Here, the  virtuality scale is 
$\tilde{Q}^2=4/r^2$), whereas BGK choose the parametric form $\tilde{Q}^2
\equiv \mu^2 = C/r^2 + \mu^2_0$ and a two-parameter initial condition for the 
gluon distribution function.

\section{Unitarity Effects in $ep$ Collisions}

This section is devoted to the study and estimate of the gluon driven
observables measured at HERA kinematical domain in the rest frame. The first
one is the inclusive structure function $F_2(x, Q^2)$, the main quantity
testing the small $x$ physics. The unitarity corrections are well established
for this observable considering Glauber-Mueller approach \cite{Ayalaepjc} as
well as its high energy asymptotics \cite{Unitbound}, namely the black disk
limit. We review these issues considering the dipole picture (rest frame),
using a more complete analysis similar to \cite{Golec,McDermott}, but mostly, 
discuss in detail  the role played by the nonperturbative physics  needed to
describe the structure function, and where in the transverse separation $r$
range it starts to be important. 

The longitudinal structure function $F_L(x,Q^2)$ is also addressed, verifying
the frame invariance in comparison with previous laboratory frame
calculations \cite{PRD59}. The longitudinal wavefunction strongly suppress
large $r$ contributions, thus selecting smaller nonperturbative
contribution in comparison with the $F_2$ case. 
Moreover, $F_L$ is one of the main observables scanning possible higher twist
corrections in the standard Operator Product Expansion (OPE) \cite{BGP}.
Therefore, a reasonable description of this quantity suggests that the
Glauber-Mueller  formalism (or similar eikonal-like approaches)  take into
account the most important contributions to the complete higher-twist
corrections at current kinematical regimes. 

The  structure function $F_2^{c\bar{c}}(x,Q^2)$ gives the charm quark
content on the proton and is directly driven by the gluon distribution.
Therefore it is a powerful observable to scan saturation effects in the small
$x$ region. However the current experimental status requires more dedicated
measurements and a better statistics. We verify a consistent description in
the rest frame corroborating the similar analysis  in the dipole models 
\cite{Donnachie,NikZoller} and in those ones considering unitarity
corrections  in the  laboratory frame  \cite{PRD59}.

\subsection{The Inclusive Structure Function $F_2(x,Q^2)$}

Now we perform estimates for the inclusive structure function in the rest
frame  considering the Glauber-Mueller dipole cross section. The expression
for $F_2$, with the explicit integration limits on photon momentum fraction
$z$ and transverse separation $r$ is, 
\begin{eqnarray}
F_2(x,Q^2)\:=\frac{Q^2}{4\,\pi^2\,\alpha_{\rm{em}}}\:\int_0^{\infty}
\!d\,^2{\bf r}\! \int_0^1 \!dz \:  \left( \vert \Psi_{T}\,(z,{\bf      
  r}) \vert ^2 + \vert \Psi_{L}\,(z,{\bf      
  r}) \vert ^2 \right)\: \sigma_{\rm{dipole}}^{GM}(x,{\bf r}^2)\,.
\label{f2dip}
\end{eqnarray} 

\begin{figure}[t]
\centerline{\psfig{figure=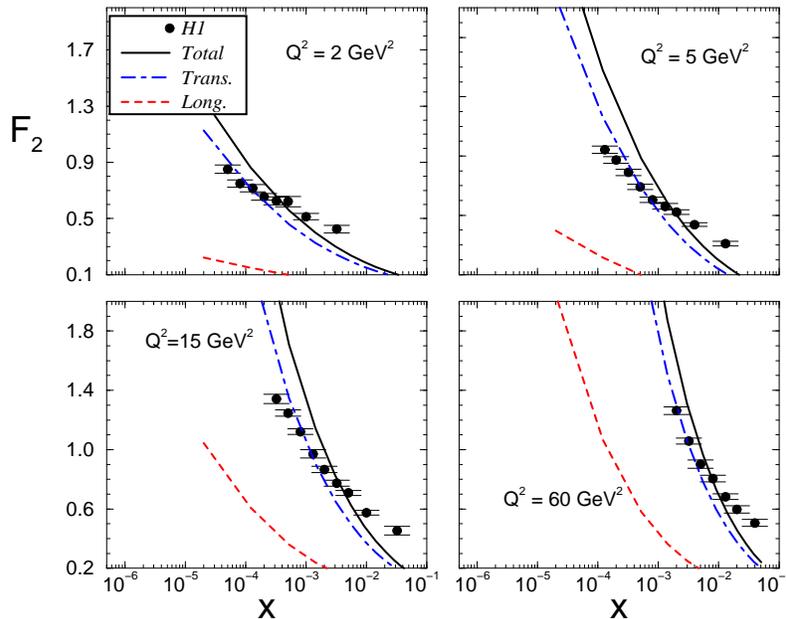,width=10cm}}       
\caption{ The Glauber-Mueller (GM) result for the $F_2(x,Q^2)$ structure
function. It is shown the transverse contribution (dot-dashed),
the longitudinal one (dashed) and total one (solid line). One considers
light quarks, target radius $R^2_A=5$ GeV$^{-2}$ and frozen
gluon distribution at large $r>r_{\rm{cut}}$. The input gluon distribution is
GRV94LO [27] and data from H1 Collaboration [30].}
\label{figf2}
\end{figure}

The notation has been introduced in the previous sections. In the Fig.
(\ref{figf2}) one shows the estimates for the  structure function for
representative virtualities $Q^2$ from the latest H1 Collaboration measurements
\cite{H1f2data}. The longitudinal and transverse contributions are shown
separately, the longitudinal one being subdominant as is well known. An
effective light quark mass ($u,d,s$ quarks) was taken, with the value
$m_q=300$ MeV, and the  target radius is considered $R^2_A=5$  GeV$^{-2}$,
in agreement with  Ref. \cite{Victorslope}. It should be stressed that
this value leads to larger saturation corrections rather than  using radius
ranging over $R^2_A\sim 8-15$ GeV$^{-2}$. The soft contribution comes from
the freezing of the gluon distribution at large transverse separation as
discussed at the previous section. The gluon distribution considered is GRV94
at leading order \cite{GRV94}, $xG^{\rm{GRV}}(x,\frac{4}{r^2})$, whose choice
has been justified in the previous section.

From the plots we verify a good agreement in the normalization, however
the slope seems quite steep. This fact is due to the modeling for the soft
contribution and it suggests that a more suitable nonperturbative input
should be taken.  Indeed, in Ref. \cite{McDermott} such a question is
addressed, claiming that the correct input is the pion-proton cross section
parametrized through the Donnachie-Landshoff pomeron. It is found that the
large transverse separations give a larger contribution at low $Q^2$, whereas 
it vanishes concerning higher virtualities. 

\begin{figure}[t]
\vspace{-2cm}
\centerline{\psfig{figure=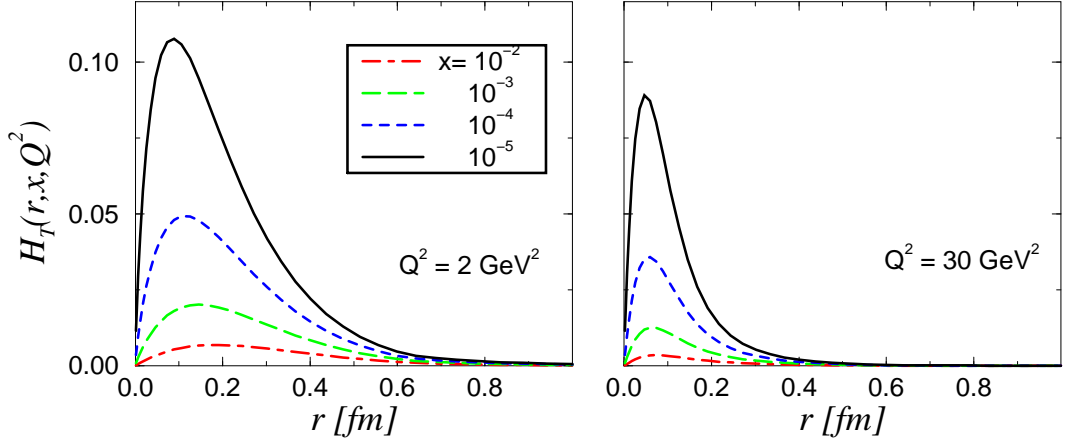,width=13cm}} 
\vspace{-2cm}      
\caption{The integrand $H_{q\bar{q}T}(r,x,Q^2)$  as a
function of $r$ for $Q^2=2$ and  $Q^2=30$ GeV$^2$ at fixed
$10^{-5}<x<10^{-2}$.} 
\label{intpartial}
\end{figure} 

To clarify this issue, we can calculate the
integrand $H_{q\bar{q}}(r,x,Q^2)$ used in the  $r$-integration for the cross
section $\sigma_{T}=\int_0^{\infty}dr H_{q\bar{q}T}(r,x,Q^2)$, that should have
significant nonperturbative content. We plot in the Fig. (\ref{intpartial})
this quantity at low virtuality $Q^2=2$ GeV$^2$ and at higher one $Q^2=30$
GeV$^2$ for momentum  fraction ranging on $10^{-4}<x<10^{-2}$, verifying that
the main contribution comes from an asymmetric peak at $r\approx 0.15$ fm
for $Q^2=2$ GeV$^2$, while it is shifted to $r\approx 0.07$ at $Q^2=30$
GeV$^2$.  In our calculation, the perturbative contribution holds up to
$r_{\rm{cut}}=0.62$ fm, therefore the region $r>r_{\rm{cut}}$ gives a non
marginal contribution to the cross section at low virtualities. Indeed, one
has found that it reaches about 10 \% at $Q^2=2$ GeV$^2$ and that when the
virtuality increases the contribution gradually vanishes. In fact, using the
most recent pdf's this situation is more critical since $r_{\rm{cut}}$ is
smaller ($Q_0^2 \sim 1-2$ GeV$^2$). This suggests that the photon piece
$|\Psi_{T}(z,r)|^2$ multiplying the dipole cross section enhances  the
$r$-integrand to smaller $r$ at high $Q^2$, corroborating a similar
conclusion  already  found in Ref. \cite{McDermott}.

\begin{figure}[t]
\centerline{\psfig{figure=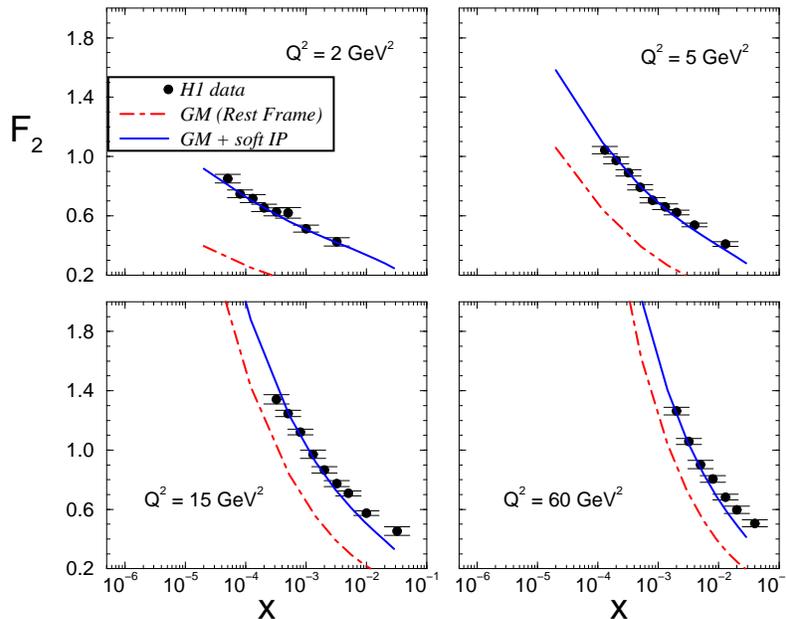,width=10cm}}       
\caption{ The Glauber-Mueller prediction for the $F_2$ structure function in
the rest frame. For sake of comparison with the results in [12]
, one uses the quark sector ($R^2_A=5$ GeV$^{-2}$, $m_q=0$) and
only the transverse wavefunction.  Radius integration $1/Q^2<r^2<1/Q_0^2$ and
soft Pomeron added (parametrizing the large pair separation  -
$F_2^{\rm{soft}}={\cal C}_{\rm{soft}}\,x^{-0.08}(1-x)^{10}$).}  
\label{f2comp} 
\end{figure}

To clarify the  role played  by the soft nonperturbative contribution to
the inclusive structure function and to verify the frame invariance of the
approach, in the Fig. (\ref{f2comp}) we plot separately the perturbative
contribution and parametrize the soft contribution introducing the
nonperturbative structure function $F_2^{\rm{soft}}={\cal
C}_{\rm{soft}}\,x^{-0.08}\,(1-x)^{10}$ \cite{Ayalaepjc}, which is added to
the perturbative one. The soft piece normalization is ${\cal
C}_{\rm{soft}}=0.22$. Such procedure is done in order to compare explicitly
the results found  in Ref. \cite{Ayalaepjc}. Accordingly, we have used just
shadowing corrections for the quark sector, taking into account only  the
transverse photon  wavefunction and zero quark mass. The integration on the
transverse separation is taken for  $1/Q^2\leq r^2 \leq 1/Q_0^2$, with
$Q_0^2=0.4$ GeV$^2$ for leading order GRV94 gluon distribution. This leads to
a residual contribution absorbed in the  the soft piece coming from transverse
separations $r^2<1/Q^2$. We considered the 
target radius being  $R_A^2=5$ GeV$^2$ (supported by \cite{Victorslope}), which
produces a correction more important   than the value $R_A^2=10$ GeV$^2$.
It is again verified that the soft contribution is important at small
virtualities and decreasing as it gets larger. In the plots, the dot-dashed
lines represent only the perturbative  calculations using the particular
approximations indicated above, and the solid lines represent the results
when we add the soft term. 

Concluding, we have a theoretical estimate, i.e. no fitting procedure,  of the
inclusive structure function $F_2(x,Q^2)$ through the Glauber-Mueller approach
for the dipole cross section, detecting a non negligible importance of a
suitable input for the large dipole size region.

\subsection{The Longitudinal Structure Function $F_L(x,Q^2)$}

As we saw in the Sec. (\ref{Sec2}), the inclusive structure function can be
expressed in terms of the cross sections $\sigma_T$ and  $\sigma_L$ for the
absortion of virtual photons transversally and longitudinally polarized,
$F_2(x,Q^2)=\frac{Q^2}{4\pi \alpha_{\rm{em}}}(\sigma_T + \sigma_L)$. At small
$x$, the longitudinal structure function is 
\begin{eqnarray}
F_L(x,Q^2)=\frac{Q^2}{4\pi \alpha_{\rm{em}}}\,\sigma_L(x,Q^2)\,.
\end{eqnarray}  
From QED, the longitudinal photons have zero helicity ($h=0$) and therefore
they have a virtual character.  In the naive parton model, the helicity
conservation for the electromagnetic vertex implies to the Callan-Gross
relation $F_2=2xF_1$ and consequently a vanishing value for the longitudinal
structure function $F_L \equiv F_2 - 2xF_1$, considering the scattering
photon-quarks (spin 1/2). From QCD theory, this quantity has a non-zero value
due to the gluon radiation, as is encoded in the Altarelli-Martinelli
expression (see discussion in \cite{PRD59})
 \begin{eqnarray}
F_L(x,Q^2)=\frac{\alpha_s(Q^2)}{2\pi}\,x^2 \int_x^1\,
\frac{dy}{y^3}\,\left[\frac{8}{3}\,F_2(x,Q^2) +
\frac{40}{9}\,y\,G(y,Q^2)\left(1-\frac{x}{y}\right)\right]\,,
\label{martinelli}
\end{eqnarray}
where $y=Q^2/sx$ is the inelasticity variable. Therefore, the structure
function $F_L$ is an auxiliary observable to detect saturation effects
(unitarity corrections) in the gluon distribution.

Experimentally, the determination of the $F_L$ is quite limited, providing 
few data points. Most recently, the H1 Collaboration has determined the
longitudinal structure function through the reduced double differential cross
section \cite{H1f2data},
\begin{eqnarray}
\sigma_r \equiv F_2(x,Q^2) - \frac{y^2}{Y_{+}}.\,F_L(x,Q^2)\,,
\end{eqnarray}
where $Y_{+}=1+(1-y)^2$. For large inelasticity, the reduced cross section
becomes $(F_2-F_L)$ and the contribution of $F_L$ is enhanced for large $y$.
The longitudinal structure function should be obtained only in the region of
large inelasticity, covered in a large range at HERA. In Ref. \cite{H1f2data},
two methods were used  to perform the extraction: (i) for large $Q^2> 10$
GeV$^2$, $F_L$ is obtained through the extrapolation method, using a NLO DGLAP
QCD fit (in the restrict kinematic range $y<0.35$ and $Q^2\leq 3.5$ GeV$^2$) to
extrapolate $F_2$ into the high $y$ region. (ii) At low $Q^2<10$ GeV$^2$, the
behavior of $F_2$ as a function of $\ln y$ is obtained using the derivative
method, based on the cross section derivative $\left( \partial
\sigma_r/\partial \ln y \right)_{Q^2}$. The data points obtained are
consistent with the previous measurements, however they are more precise and
lying into a broader kinematical range. Therefore, in the following we use
only the new data points to analyse.

\begin{figure}[t]
\centerline{\psfig{figure=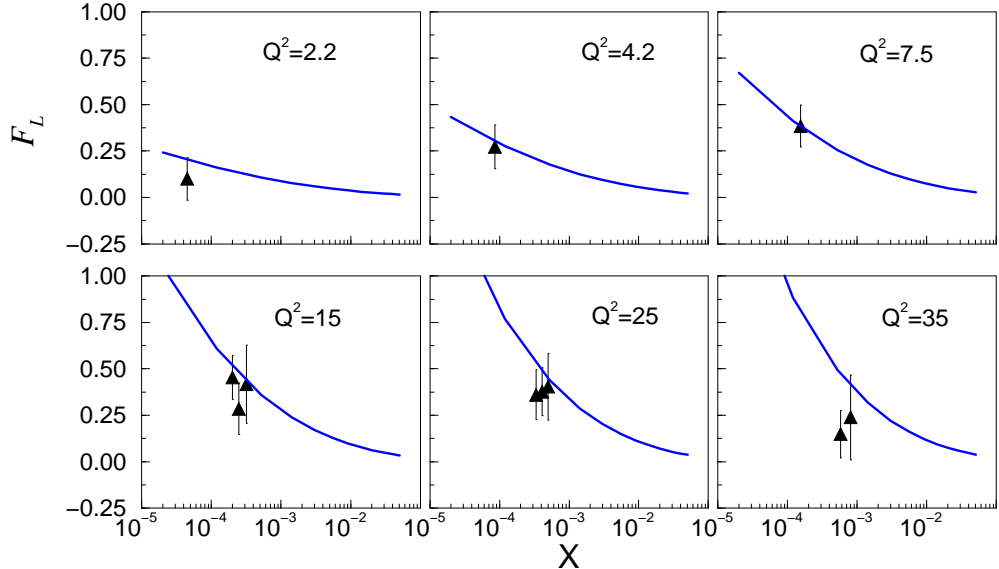,width=12cm}}       
\caption{ The Glauber-Mueller estimates for the $F_L$ structure function.
One uses light quarks ($m_{q}=300$ MeV), target size $R^2_A=5$ GeV$^{-2}$ and
frozen gluon distibution at large $r$.  Data from H1 Collaboration [30].}
\label{flrframe}
\end{figure}
In Fig. (\ref{flrframe}) we present the estimates for the $F_L$ structure
function, in representative virtualities as a function of $x$.   For the
calculations, it was considered light quarks ($u,\,d,\,s$) with effective mass
$m_{q}=300$ MeV and  the target radius $R^2_A=5$ GeV$^{-2}$. The large $r$
region is considered by the freezening of the gluon distribution at that
region. Our expression for the observable is then, \begin{eqnarray}
F_L(x,Q^2)\:=\frac{Q^2}{4\,\pi^2\,\alpha_{\rm{em}}}\:\int_0^{\infty}
\!d\,^2{\bf r}\! \int_0^1 \!dz \:   \vert \Psi_{L}\,(z,{\bf         r}) \vert
^2 \: \sigma_{\rm{dipole}}^{GM}(x,{\bf r}^2)\,. \end{eqnarray} 

The behavior is in agreement with the experimental result, either in shape
as in normalization. A better description can be obtained by fine tunning  the
target size or the considered gluon distribution function, however it should be
stressed that the present prediction is parameter-free and determined using the
dipole picture taking into account unitarity (saturation) effects in the
effective dipole cross section.  We verify that the rest frame calculation,
taking into account the dipole degrees of freedom and unitarity effects
produces similar conclusions to those ones using the Breit system.
For instance, in a previous work \cite{PRD59}, the unitarity corrections to the
longitudinal structure function were estimated in the laboratory frame
considering the Eq. (\ref{martinelli}), with unitarized expressions for $F_2$
and $xG(x,Q^2)$, obtaining that the  expected
corrections reach  to   $ 70$ \% as $\ln (1/x)=15$, namely on the
kinematical sector of the upcoming THERA project.

The higher twist corrections to the longitudinal structure function
have been pointed out. For instance, Bartels et al. \cite{BGP} have calculated
numerically the twist-four correction obtaining that they are large for  $F_T$
and $F_L$, although with opposite signs. This fact leads to  remaining
small effects to the inclusive structure function by almost complete
cancellation between those contributions. The higher twist content is analyzed
considering the model \cite{Golec} as initial condition. 

Concerning $F_L$, it was found that the
twist-four correction is large and has negative signal, concluding that a
leading twist analysis of $F_L$  is unreliable for high $Q^2$ and not too small
$x$. The results are in agreement with the simple parametrization for higher
twist (HT) studied by the MRST group in Ref. \cite{MRSTHT}, where
$F_2^{HT}(x,Q^2)=F_2^{LT}(x,Q^2)(1+\frac{D_2^{HT}(x)}{Q^2})$. The second term
would parametrize the higher twist content. In our case, the unitarity
corrections provide an important amount of higher twist content, namely it
takes into account some of the several graphs determining the 
twist expansion (for recent discussions in these issues, see \cite{Lev2}).

\subsection{The Charm Structure Function $F_2^{c\bar{c}}(x,Q^2)$}

In perturbative QCD, the heavy quark production in electron-proton
interaction occurs basicaly through photon-gluon fusion, in which the emitted
photon interacts with a gluon from the proton generating a quark-antiquark
pair. Therefore, the heavy quark production allows to determine the gluon
distribution and the amount of unitarity  (saturation) effects for the
observable. In particular, charmed mesons have been measured at deep-inelastic
at HERA and the  corresponding structure function $F_2^{c\bar{c}}(x,Q^2)$ is
defined from the differential cross section for the $c\bar{c}$ pair production,
\begin{eqnarray}
\frac{d^2\,\sigma^{c \bar{c} }}{dx \, dQ^2 }=\frac{2\pi
\alpha_{\rm{em}}}{x\,Q^4} \left[\,1+(1-y)^2\,\right]\,F_2^{c\bar{c}}(x,Q^2)\,,
\end{eqnarray}
with $y$ being the inelasticity variable. In the laboratory frame, the dominant
mechanism is the boson-gluon fusion $\gamma^* G \rightarrow c\bar{c}$. Hence,
the charm structure function is directly driven by the gluon distribution and
provides constraints for the gluonic function. In leading order (LO), it is
written as \cite{PRD59},
\begin{eqnarray}
F_2(x,Q^2,m_c^2)=\frac{4\,\alpha_s(\mu_F^2)}{9\pi}\,\int_{a_cx}^{1}\,\frac{dy}{y}\,C^{c}_{g,\, 2} \left(\frac{x}{y}, 
\,\frac{m^2_c}{Q^2}\right)\, xG(y,\mu_F^2) \,,
\label{f2ccms}
\end{eqnarray}
where $a_c=(1+\frac{m^2_c}{Q^2})$. The mass factorization scale lies in the
range $m^2_c\leq  \mu_F^2 \leq 4\,(Q^2 + 4\,m_c^2)$. Such a scale introduces
an uncertainty of about 10 \%, and an additional source of uncertainty is the
charm mass, in general ranging on $1.2 \leq m_c \leq 1.7$ GeV.  The standard
QCD coefficient function is labeled by $C^c_{g,\,2}\left( z,\frac{m^2_c}{Q^2}
\right)$.

Experimentally, the latest measurements of the charm structure function are
obtained by measuring mesons $D^{*\,\pm}$ production \cite{ZEUScharm}. From the
theoretical input, it was used NLO coefficient functions, considering charm
mass $m_c=1.4$ GeV and factorization-normalization scale
$\mu_F=\sqrt{Q^2+4\,m^2_c}$. The function $F_2^{c\bar{c}}(x,Q^2)$  shows an
increase with decreasing $x$ at constant values of $Q^2$, whereas the rise
becomes more intense at higher virtualities. The data are consistent with the
NLO DGLAP calculations. Concerning the ratio
$R^{c\bar{c}}=F_2^{c\bar{c}}/F_2$, the charm contribution to $F_2$ grows
steeply as $x$ diminishes. It contributes less than 10\% at low $Q^2$ and
reaches to about 30 \% for $Q^2>120$ GeV$^2$ \cite{ZEUScharm}. 

Once more the color dipole picture will provide a quite simple description
for the charm structure function in a factorized way. Now, the Glauber-Mueller
dipole cross section is weighted by the photon wavefunction constituted  by a
$c\bar{c}$ pair with mass $m_c$.  Our expression for the charmed contribution
in deep inelastic is thus  written as  \begin{eqnarray}
F_2^{c\bar{c}}(x,Q^2)\:=\frac{Q^2}{4\,\pi^2\,\alpha_{\rm{em}}}\:\int_0^{\infty} \!d\,^2{\bf r}\! \int_0^1 \!dz \:  \left( \vert \Psi^{c\bar{c}}_{T}\,(z,{\bf         r}) \vert ^2 + \vert \Psi^{c\bar{c}}_{L}\,(z,{\bf         r}) \vert ^2 \right)\: \sigma_{\rm{dipole}}^{GM}(x,{\bf r}^2) \end{eqnarray}  where $\vert \Psi^{c\bar{c}}_{T,\,L}\,(z,{\bf      
  r}) \vert ^2$ is the probability to find in the photon the $c\bar{c}$ color
dipole with the charmed quark carrying fraction $z$ of the photon's light-cone
momentum with $T,\,L$ polarizations. For the correspondent wavefunctions, the
quark mass in Eqs. (\ref{eq9},\ref{eq10}) should be substituted by the charm
quark  mass $m_c$. Here, we should take care of the connection between the
Regge parameter $x=(W^2+Q^2)/(Q^2+ 4\,m^2_q)$ and the Bjorken variable
$x_{\rm{Bj}}$. For calculations with the light quarks these variables are
equivalent, however for heavier quarks the correct relation is
\cite{NikZoller},  \begin{eqnarray}
x_{\rm{Bj}}=x\,\left( \frac{Q^2}{Q^2 + 4\,m_c^2} \right)\,.
\label{xbj}
\end{eqnarray}

\begin{figure}[t]
\vspace{-2cm}
\centerline{\psfig{figure=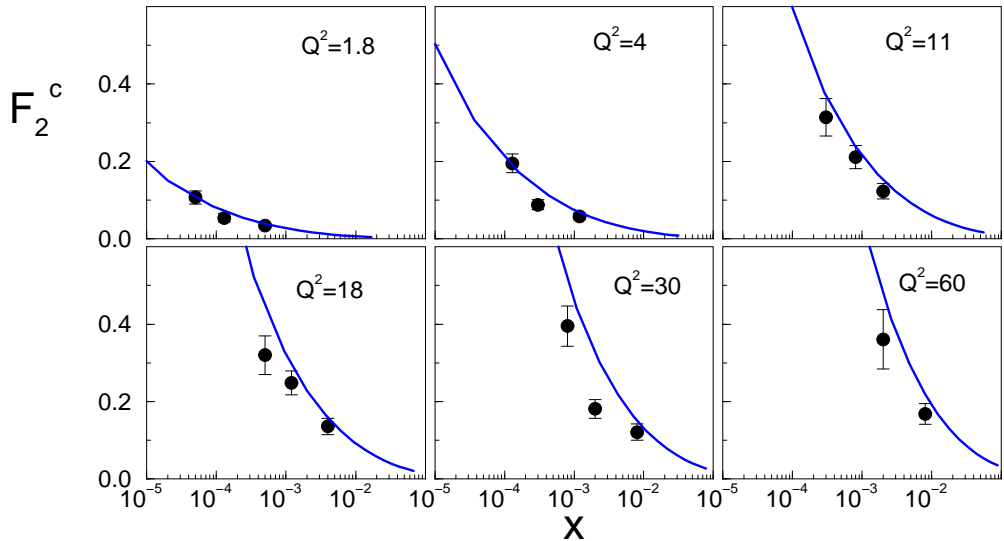,width=12cm}}       
\caption{ The Glauber-Mueller result for the $F_2^{c\bar{c}}$ structure
function as a function of Bjorken variable $x$ at fixed virtualities (in
GeV$^2$). One uses charm mass $m_{c}=1.5$ GeV, target size $R^2_A=5$ GeV$^{-2}$
and frozen gluon distibution at large $r$.  Data from ZEUS Collaboration [33]
 (statistical errors only).}  \label{f2cfig}
 \end{figure}

In Fig. (\ref{f2cfig}) we show the estimates for the charm structure function
as a function of $x_{\rm{Bj}}$ [Eq. (\ref{xbj})] for representative
virtualities. In our calculations, it was used the charm mass $m_c=1.5$ GeV,
the target size $R^2_A=5$ GeV$^{-2}$ and the frozen gluon distibution at large
$r$. We have verified a small soft contribution as in the $F_L$ case,
decreasing as the virtuality rises. There is a slight sensitivity to the value
of the charm mass, increasing the overall normalization as $m_c$ diminishes.
Such a feature  suggests that  the charm mass is a hard scale  suppressing  the
non-perturbative contribution to the corresponding cross section, which  is
in agreement with the recent BFKL color dipole calculations of Nikolaev-Zoller
\cite{NikZoller} and  Donnachie-Dosch \cite{Donnachie}.

Regarding the Breit system description, in Ref. \cite{PRD59} it  was found
strong corrections to the charm structure function, which are larger than
those of the $F_2$ ones making use of expression (\ref{f2ccms}). Considering
the ratio $R_2^c=F_2^{c\,\rm{GM}}(x,Q^2)/F_2^{c\,\rm{DGLAP}}(x,Q^2)$, the
corrections predicted by the Glauber-Mueller approach would reach  to 62 \%
at values of $\ln (1/x) \approx 15$ (THERA region). Then, an important result
is a large deviation of the standard DGLAP expectations at small $x$ for the
ratio $R^{c\bar{c}}=F_2^{c\bar{c}}/F_2$ due to the saturation phenomena
(unitarization). With our calculation one verifies that it is obtained  a good
description of data in both reference systems, suggesting a consistent
estimation of the unitarity effects for this quantity.

\section{Conclusions}

We study the dipole picture for the description of deep inelastic
scattering, focusing on observables driven directly by the gluon
distribution. Starting from the dipole cross section provided by the
Glauber-Mueller approach in QCD, we perform estimates for the inclusive
structure function $F_2$, the longitudinal function $F_L$ and  the charm
structure function on the proton $F_2^{c\bar{c}}$.

For each of the observables discussed, we obtain theoretical estimates, in the
rest frame,  without further fitting procedure, in good agreement with the
updated data from HERA. The resulting calculations corroborate a quite
consistent picture for the unitarity corrections from the Glauber-Mueller
approach in both Breit and rest reference systems. In the laboratory frame the
unitarity effects are connected with the gluon distribution function, whereas
in the color dipole framework the basic block is the dipole cross section
which is corrected considering saturation effects. 

The small transverse separation $r$ region is dominated by the leading log
DGLAP formalism, with the additional ingredient of unitarization
phenomenon as the momentum fraction acquires quite small values. Such
corrections are associated with the taming of the gluon distribution in the
very small $x$ region, in general named saturation regime. However, it should
be stressed that the Glauber-Mueller approach and similar eikonal-like models
provide  a logarithmic $\ln(1/x)$ asymptotic behavior for the inclusive
structure function and gluon distribution, instead of a constant value for 
asymptotic energies.

The large transverse separation is described by non-perturbative aspects of
QCD. Since this domain is not well determined at the moment, some  modelling
of the soft region is needed. In this work we choose the  ans\"{a}tz in which  
the gluon distribution is frozen for virtualities above a cut
radius $r^2>r^2_{\rm{cut}}$ , which corresponds to the region $Q^2<Q^2_0$. A convenient  choice for the  gluon pdf in order to cover the
widest possible kinematical window  diminishes the
uncertainty coming from the soft sector. The
most appropriated input is the GRV94 parametrization, where
$r_{\rm{cut}}=0.6$ fm is found, whereas it can  take values
$r_{\rm{cut}}=0.4-0.5$ fm for the more recent pdf's. Throughout the paper we
used the target size $R_A^2=5$ GeV$^{-2}$, which corresponds to strong
unitarity corrections.

When performing a comparison with the phenomenological model GBW, we have
found that the Glauber-Mueller approach underestimate the dipole cross section
from GBW at not small $x\geq 10^{-3}$. Instead, for very low $x\leq 10^{-4}$
the Glauber-Mueller overestimates GBW due to the strong increasing of the gluon
function in this region. Concerning the saturation model with DGLAP evolution
(BGK), it is a particular case of the GM approach when considering central
scattering $b=0$. Despite that the BGK model matches DGLAP evolution,
Glauber-Mueller describes more properly the realistic impact parameter
dependence of the process. Moreover, in GM the extension to the nuclear case
is built in.

When considering the structure function $F_2$, we have found that it is
dominated by small transverse distances contributions. However, a
non-negligible content from the soft sector is present. Moreover, the photon
wavefunctions enhances the dipole cross section into smaller dipole sizes,
since the  weight function selects smaller $r$ as the virtuality $Q^2$
diminishes. Our estimates here are parameter-free, however a fine tunning of
the parameters can improve the data description. Furthermore, we notice that
in calculations from  \cite{Ayala}, only the aligned jet dipole configuration
$z$, $(1-z) \approx 0$ (and only transverse contribution) is considered,
whereas we take into account all configurations, including the symmetric ones.
Thus, all dipole sizes, even those from the non-perturbative region are
included in our results.

Concerning $F_L$, the estimates are consistent with the previous calculations
in the Breit system and are in good agreement with data. A remarkable feature
is that the Glauber-Mueller approach in the color dipole framework gives
important higher twist contributions to the leading twist calculation in a
simple way. As is well known, $F_L$ is the main quantity to study the expected
higher twist effects in low virtualities. 

The function $F_2^{c\bar{c}}$ is directly dependent of the gluon distribution
and  important unitarity corrections had been found when considering the
Breit frame. Here, we verify consistent results in the  rest frame in
comparison with the previous ones in the fast proton system. We verified that
the charm mass suppressed soft contributions in comparison with the $F_2$ case,
and the results present a slight dependence with the specific value of $m_c$.

In conclusion, the Glauber-Mueller approach provides a well stablished
formalism to take into account the unitarity effects. It allows to estimate
the higher twist contributions to relavant observables in a
simplified way, while  matching  DGLAP evolution equation at Born level
and including the impact parameter dependence properly.

\section*{Acknowledgments}

MVTM acknowledges  Martin
McDermott (Liverpool University-UK), Igor Ivanov (IKP-Forschungszentrum
Juelich, Germany) and   Victor Gon\c{c}alves (IFM-UFPel, Brazil) for useful
enlightenements.  This work was partially financed by CNPq, Brazil.

\end{document}